\documentclass[pra,aps,twocolumn,10pt,showpacs,nofootinbib]{revtex4-1}
\usepackage[utf8]{inputenc}
\usepackage[T1]{fontenc}
\usepackage{amsmath}
\usepackage{amsfonts}
\usepackage{amssymb}
\usepackage{mathcmd}
\usepackage{graphicx} 

\usepackage{braket}
\usepackage{bm}
\usepackage{mathrsfs}
\usepackage{stmaryrd}

\usepackage{tikz}
\usepackage{rotating} 
\usepackage{float}

\newcommand{\mi}{\mathrm{i}}
\newcommand{\me}{\mathrm{e}}
\newcommand{\md}{\mathrm{d}}

\begin{document}

\title{Geometrically robust linear optics from non-Abelian geometric phases}
\author{Julien Pinske}
\author{Stefan Scheel}
\email{stefan.scheel@uni-rostock.de}
\affiliation{Institut f\"ur Physik, Universit\"at Rostock, 
Albert-Einstein--Stra{\ss}e 23-24, D-18059 Rostock, Germany}

\date{\today}

\begin{abstract}
We construct a unified operator framework for quantum holonomies 
generated from bosonic systems. For a system whose Hamiltonian is 
bilinear in the creation and annihilation operators, we find a holonomy group 
determined only by a set of selected orthonormal modes obeying a stronger 
version of the adiabatic theorem. This photon-number independent description 
offers deeper insight as well as a computational advantage when compared to the 
standard formalism on geometric phases. In particular, a strong analogy between 
quantum holonomies and linear optical networks can be drawn. This relation 
provides an explicit recipe how any linear optical quantum computation can be 
made geometrically robust in terms of adiabatic or nonadiabatic geometric 
phases. 
\end{abstract}
\maketitle


\section{Introduction}
Recent years have witnessed an increased interest in the notions of Abelian 
and non-Abelian geometric phases (quantum holonomies) beyond the gauge theories 
of elementary particles. The experimental and theoretical study of such phase 
factors is relevant to the simulation of lattice gauge theories 
\cite{Zoller,Cirac}, the understanding of symmetry groups 
\cite{Goldman,Spielman}, as well as much of modern mathematics \cite{Chern}. 
Moreover, their topological properties give rise to quantum evolutions that are 
inherently fault-tolerant and might therefore be highly desirable assets for 
quantum information processing \cite{HQC,Sjoeqvist,Tong}.

These purely geometric transformations arise when a state vector is parallelly 
transported along a closed loop in a suitable subspace. For adiabatic 
holonomies this subspace is usually some (non-)degenerate ground-state subspace  \cite{Berry,Wilzeck}. In the case of a 
nonadiabatic quantum holonomy, evolution takes place in a subspace on which the 
mean energy vanishes, thus making the evolution purely geometric 
\cite{Anandan}. Physical implementations rely, for instance, on atomic 
transitions \cite{Recati,Feng}, the manipulation of trapped ions 
\cite{Duan,Guo} or superconducting qubits \cite{Abdumal,Xu}, as well as 
controlled guiding of coherent light \cite{Teuber,Conti} or photons 
\cite{Pachos}. The latter implementation is of particular interest as the 
dimension of the relevant subspace increases with a rising number of photons 
\cite{Pinske2020}. This hints at a formulation of geometric phases that is 
independent of the overall particle number to which the system is subjected. 

In this article, we present an operator-based formalism for the unified treatment of adiabatic and nonadiabatic quantum holonomies in terms of a 
holonomic Heisenberg picture that generalises the concept of an adiabatic 
Heisenberg picture first introduced in Ref.~\cite{Brihaye}. It bears some 
resemblance to (projective adiabatic) elimination procedures \cite{Sanz} and
relies on the usage of effective Hamiltonians \cite{Biswas,Fujikawa} to 
describe the geometric phase of a system. The holonomic Heisenberg picture 
developed here 
employs a generalised
parallel transport condition that imposes the disappearance of Hamiltonian dynamics on a specific set of solutions. Remarkably, in a linear optical setting we find 
a quantum holonomy on the level of superoperators that only depends on a set of 
selected orthonormal modes of the underlying system. Subsequently, this 
eliminates the need for an explicit calculation of projectors onto the 
relevant subspace, which becomes quickly unfeasible for large bosonic systems. 
Besides offering a major computational advantage over the standard formalism, 
it also provides deeper insight into the emergence of geometric phases in 
second quantisation. 
From our general argument it follows that linear optical setups based on 
photonic holonomies can be described equally by a holonomic scattering matrix in 
analogy to the Reck-Zeilinger scheme \cite{Reck}. Thus, it is possible to make 
any linear optical computation geometrically robust by referring to auxiliary 
modes while undergoing cyclic evolution. We benchmark our findings by 
studying a number of examples which are important to modern quantum optics. 

\section{Operator formulation of quantum holonomies}
The treatment of bosonic many-particle 
systems can be pursued elegantly by referring to a Fock representation.
Here, the Hilbert space of a system consisting of $M$ modes is the 
Fock space $\mathscr{H}=\bigotimes_{k=1}^{M}\mathscr{H}_k$ defined as a finite 
tensor product of single-mode Fock spaces 
$\mathscr{H}_k=\mathrm{Span}\{\ket{n}_k\,|\, n\in\mathbb{N}_0\}$. Starting from 
the multi-mode vacuum state $\ket{\bm{0}}=\ket{0,\dots,0}$, the Fock space 
$\mathscr{H}$ can be viewed as being generated from the (unital $*$-)algebra 
$\mathscr{A}=\bigotimes_{k=1}^{M}\mathscr{A}_k$ containing analytic functions of 
creation and annihilation operators $\{\hat{a}_{k},\hat{a}^{\dagger}_{k}\}_k$ 
satisfying the canonical commutation relations 
$[\hat{a}_{j},\hat{a}_{k}^{\dagger}]=\delta_{jk}$ and 
$[\hat{a}_{j},\hat{a}_{k}]=[\hat{a}_{j}^{\dagger},\hat{a}_{k}^{\dagger}]=0$.
We express their relation as $\mathscr{H}\cong\mathscr{A}\ket{\bm{0}}$, which 
says that any multi-mode state $\ket{\Psi}\in \mathscr{H}$ can be expanded 
as $\ket{\Psi}=\hat{F}(\hat{a}_{1}^{\dagger},\dots,\hat{a}_{M}^{\dagger})
\ket{\bm{0}}$, where 
\[
\begin{split}
\hat{F}=\sum_{p_1,\dots,p_M\in\mathbb{N}_0}c_{p_1,\dots, p_M}
\frac{\big(\hat{a}_{1}^\dagger\big)^{p_{1}}}{\sqrt{p_1!}}\dots 
\frac{\big(\hat{a}_{M}^\dagger\big)^{p_{M}}}{\sqrt{p_M!}},
\end{split}
\]
with $\sum_{p_1,\dots,p_M}|c_{p_1,\dots, p_M}|^2=1$.

Let us consider a time-dependent Hamiltonian $\hat{H}(t)$ 
belonging to $\mathscr{A}$. While the general evolution of a quantum state $\ket{\Psi(t)}$ is subject to Schrödinger's equation $\mi\frac{\md}{\md t}\ket{\Psi}=\hat{H}\ket{\Psi}$,
a quantum holonomy transforms the state in a manner that is independent of the spectrum of $\hat{H}(t)$ or runtime $T$. Certainly not any solution to Schrödinger's equation will be of such nature, but only those belonging to a subset of solutions $\mathscr{K}(t)=\{\ket{\eta_{j}(t)}\}_j$. As these states evolve purely geometrically, they shall satisfy  $\braket{\eta_j|\dot{\eta}_k}=0$, thus being independent of the systems dynamics. More formally, we will summarise the latter as $\braket{\frac{\md}{\md t}}_{\mathscr{K}}=0$ being a condition for the parallel transport of a quantum state. 
Expanding the solutions in $\mathscr{K}$ in terms 
of an orthonormal basis $\{\ket{\psi_j}\}_{j}$ of a subspace $\mathscr{H}_{\psi}$ as 
$\ket{\eta_k(t)}=\sum_{j}U_{jk}(t)\ket{\psi_j(t)}$ with initial condition 
$\ket{\eta_k(0)}=\ket{\psi_k(0)}$, and employing the law for parallel transport 
leads to the first-order differential equation $(\hat{U}^{-1} 
\dot{\hat{U}})_{jk}=(\hat{A}_{t})_{kj}=\braket{\psi_j|\dot{\psi}_k}$. A formal 
solution is given by the time-ordered matrix exponential 
$\hat{U}(T)=\mathcal{T}\me^{\int_{0}^{T}\hat{A}_{t}\md t}$. In this context, and throughout the article, $\mathscr{H}_\psi$ plays the role of a geometrically protected subspace. Here we are not interested in arbitrary evolutions but in those that represent 
a loop $\gamma(t)$ through the subspace, that is $\mathscr{H}_\psi(0)=\mathscr{H}_\psi(T)$. The unitary is then given by a 
path-ordered matrix exponential
$\hat{U}(\gamma)=\mathcal{P}\me^{\oint_\gamma 
		\hat{A}}$
known as a quantum holonomy \cite{Wilzeck,Anandan}. 
The connection might then be expressed through its 
anti-Hermitian components $(\hat{A})_{jk}=\bra{\psi_{k}}\md\ket{\psi_j}$ 
mediating parallel transport along the loop $\gamma$. 
In order for expectation values to coincide on the subspace 
$\mathscr{H}_{\psi}$, any mode $\hat{a}_k$ in $\mathscr{A}$ must evolve according to 
$\hat{a}_k\mapsto \hat{U}^\dagger(\gamma)\hat{a}_k\hat{U}(\gamma)$. This implies 
that the Heisenberg equation of motion for a purely holonomic evolution reads
\begin{equation}
	\label{eq:Op}
\braket{\dot{\hat{a}}_k}_{\mathscr{H}_{\psi}}=\braket{[\hat{a}_k,\hat{A}_{t}(\gamma)]}_{\mathscr{H}_{\psi}}+\braket{\partial_t\hat{a}_k}_{\mathscr{H}_{\psi}},
\end{equation}
which means that the 
generator of dynamics is now given by a parallel transport map 
$\hat{A}_{t}(\gamma)=\hat{U}^\dagger(\gamma)\hat{A}_{t}\hat{U}(\gamma)$ instead of a 
general Hamiltonian. Having determined the evolution of modes $\hat{a}_k$ from Eq.~(\ref{eq:Op}), then characterises adiabatic changes of any function $\hat{F}(\hat{a}_k,\hat{a}_k^\dagger)$ in the algebra $\mathscr{A}$.

For concreteness, if we consider $\mathscr{K}$ to be the set 
of adiabatic solutions, the states $\{\ket{\psi_j(t)}\}_j$ form an orthonormal 
basis for the degenerate ground-state subspace $\mathscr{H}_\psi=\mathscr{H}_0$ of the Hamiltonian $\hat{H}(t)$. 
When traversing this loop slowly, by which we mean the $\ket{\psi_j(t)}$ change only gradually when compared to the energy gap between $\mathscr{H}_0$ and the excited states of the system, we recover the adiabatic Heisenberg picture 
\cite{Brihaye}. To be more precise, note that adiabatic solutions only approximate the evolving state governerd by Schrödingers equation up to first order of 
$1/(T\Delta\varepsilon)$, where $\Delta\varepsilon$ is the energy gap between the ground and excited states. As an elementary example consider the adiabatic propagation through the zero-eigenvalue eigenspace $\mathscr{H}_0$ of a nonlinear Kerr medium \cite{PachosOpt} 
$\hat{W}(\alpha,\xi)\hat{H}_0\hat{W}^\dagger(\alpha,\xi)$, with 
	$\hat{H}_0=\hat{a}^\dagger \hat{a}(\hat{a}^\dagger \hat{a}-1)$ and 
	$\hat{W}(\alpha,\xi)=\me^{\alpha 
		\hat{a}^\dagger-\alpha^*\hat{a}}\me^{\frac{\xi^*}{2}\hat{a}^2-\frac{\xi}{2}(\hat
		{a}^\dagger)^2}$ describing the combined process of coherent displacement and single-mode squeezing. The action of $\hat{a}$ onto $\mathscr{H}_0$ is governed by the connection 
	$\hat{A}_t=\hat{\Pi}_0\hat{W}^\dagger \partial_t\hat{W}\hat{\Pi}_0$, with 
$\hat{\Pi}_0=\ket{0}\bra{0}+\ket{1}\bra{1}$. Evaluating Eq.~(\ref{eq:Op}) leads to nonlinear equations of adiabatic motion 
\[\braket{\dot{\hat{a}}}_{\mathscr{H}_0}=(\dot{\alpha}^*\alpha-\dot{\alpha}\alpha^*)\braket{\hat{a}}_{\mathscr{H}_0}
+\dot{\alpha}(\mu-\nu^*)\braket{\hat{a}^\dagger\hat{a}}_{\mathscr{H}_0}-\mathrm{c.c.},
\]
where $\mu=\cosh|\xi|$ and $\nu=\me^{\mi\,\mathrm{arg}(\xi)}\sinh|\xi|$. The emergence of nonlinear equations of motion is a generic feature of the operator description of parallel transport, both adiabatic and nonadiabatic. This is due to the connection $\hat{A}=\hat{\Pi}_\psi\md\hat{\Pi}_\psi$ requiring the computation of subspace projectors $\hat{\Pi}_\psi$ onto 
$\mathscr{H}_\psi$. Due to the generally highly nonlinear form of these 
projectors in terms of bosonic modes, the computation of quantum holonomies 
can be an extremely challenging task.

For completeness, note that the above argument extends to any eigenspace with energy 
$\varepsilon_n(t)$. When level-crossing is neglected, i.e. if $n\neq m$, then $\varepsilon_n(t)\neq\varepsilon_m(t)$ for all $t\in[0,T]$, and as a result the degeneracy of each energy level does not change. 
The overall time evolution of $\hat{a}_k$
under the adiabatic assumption (long runtime $T$) is then determined by the composite unitary $\bigoplus_n 
\me^{\mi\int_{0}^{T}\varepsilon_n(t)\md t}\hat{U}_n(\gamma)$
with $\hat{U}_n(\gamma)$
being the holonomy acting on the $n$th eigenspace of 
the system Hamiltonian, and $\int_{0}^{T}\varepsilon_n(t)\md t$ 
accounting for dynamical contributions. Note that at first glance the parallel 
transport condition $\braket{\frac{\md}{\md t}}_{\mathscr{K}}=0$ might be 
violated in an eigenspace with $\varepsilon_n(t)\neq 0$. However, this can 
always be accounted for by multiplying the solutions in $\mathscr{K}$ with the 
dynamical phase $\me^{\mi\int_{0}^{T}\varepsilon_n(t)\md t}$. Therefore, strictly speaking, the composite unitary is not a fully geometric quantity but has dynamical contributions due to these relative (energy-dependent) phase factors. 

\section{Linear quantum optics}
The general concepts described thus far can, in principle, be applied to any bosonic system. In the following, we will show that in a linear optical setting, that is the Hamiltonian $\hat{H}(t)$ is bilinear in the creation and annihilation operators, certain symmetries arise that offer a deeper insight into the emergence of geometric phases.
Consider a system of $M$ bosonic modes that interact 
according to such a bilinear Hamiltonian. Suppose further that there is a set of orthonormal modes $\{\hat{\Psi}_j(t)\}_{j=1,}^{K<M}$ whose excitations (action on $\ket{\bm{0}}$) span a subspace $\mathscr{H}_\psi=\{\ket{\psi_l}, l\in\mathbb{N}\}$ on which the mean energy of the Hamiltonian $\hat{H}$ vanishes. In Appendix~\ref{app:quant} we 
show that the second-quantisation formulation of the condition $\braket{\frac{\md}{\md t}}_{\mathscr{K}}=0$ is given as
$[\hat{\Psi}_j,[\hat{H},\hat{\Psi}_k^\dagger]]=0$.
This implies $\bra{\psi_l}\hat{H}\ket{\psi_m}=0$ for any $l,m\in\mathbb{N}$, thus ensuring 
that the evolution is indeed of purely geometric origin. In the same way, we can view the solutions in $\mathscr{K}$ as being created by operators $\hat{\eta}_k^\dagger(t)$, $k=1,\dots,K$, which then must satisfy the Heisenberg equation of motion. With the ansatz 
$\hat{\eta}_k^\dagger(t)=\sum_{j}\mathcal{U}_{jk}(t)\hat{\Psi}_j^\dagger(t)$ and 
the condition for parallel transport, this yields 
\begin{equation}
 \label{eq:AQH}
 0=[\hat{\eta}_j,\dot{\hat{\eta}}_k^\dagger]=\sum_{l=1}^{K} 
\mathcal{U}_{lj}^*\dot{\mathcal{U}}_{lk}
+\sum_{l,m=1}^{K}\mathcal{U}_{lj}^*\mathcal{U}_{mk}
\left(\mathcal{A}_{t}\right)_{ml},
\end{equation}
where we introduced the operator-valued connection 
$\left(\mathcal{A}_{t}\right)_{jk}=[\hat{\Psi}_k,\dot{\hat{\Psi}}_j^\dagger]=[\hat{\Psi}_k,\partial_t\hat{\Psi}_j^\dagger]$. If we now consider a cyclic evolution of the system, i.e. for $j=1,\dots,K$ we have $\hat{\Psi}_j^\dagger(0)=\hat{\Psi}_j^\dagger(T)$ resembling a loop $\gamma$, the solution to Eq.~(\ref{eq:AQH}) is formally 
given by the path-ordered integral superoperator 
$\mathcal{U}_\gamma=\mathcal{P}\mathrm{exp}\oint_{\gamma}\mathcal{A}$.
Now, the time evolution of a mode $\hat{\eta}_k^\dagger$ is given by 
the mapping $\hat{\eta}_k^\dagger(T)=\mathcal{U}_{\gamma}[\hat{\eta}_k^\dagger(0)]$.

Strikingly, in this formulation one avoids the usage of projectors onto the 
relevant subspace altogether, thereby drastically simplifying the computational 
effort needed to determine the geometric evolution. Starting 
at a point where
$\hat{\eta}_k^\dagger(0)=\hat{a}^\dagger_k$, it becomes evident that 
$\mathcal{U}_\gamma$ can be viewed as the scattering matrix of the linear 
optical network being restricted to purely geometric evolutions of the bosonic 
modes. Note that, because only $K<M$ modes are relevant to the final output of the network, there are $M-K$ remaining auxiliary modes that act as mediators for a purely geometric evolution.
This result can be related to the standard formalism on geometric 
phases \cite{Wilzeck,Anandan} by noting 
that $\mathcal{U}_{\gamma}[\hat{a}_k^\dagger]
=\hat{U}^\dagger(\gamma)\hat{a}_k^\dagger\hat{U}(\gamma)$, with 
$\hat{U}(\gamma)=\mathcal{P}\me^{\oint_{\gamma}\hat{A}}$ being the more 
familiar form of the holonomy. The associated connection can be obtained from 
$\hat{A}_\mu=\sum_{j,k}(\mathcal{A}_\mu)_{jk}\hat{a}^\dagger_j\hat{a}_k$ being 
bilinear in the creation and annihilation operators. In contrast to a nonlinear 
optical setting, here the projection onto the relevant subspace is incorporated 
implicitly into the connection, thus providing an elegant photon-number 
independent description.

\subsection{Geometric picture of operator holonomies} 
\begin{figure}[h]
\begin{tikzpicture}
\node at (0,0) {\includegraphics[width=5.5cm]{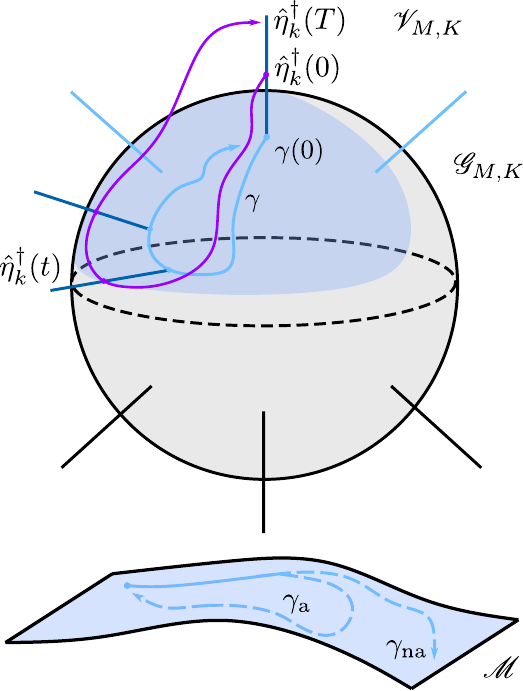}}; 
\end{tikzpicture}
\caption{\label{fig:bundle} The horizontal lift $\{\hat{\eta}^\dagger_k(t)\}_k$ 
moves along the fibres over the loop $\gamma$ (dark blue spikes). The 
difference between $\hat{\eta}^\dagger_k(0)$ and $\hat{\eta}^\dagger_k(T)$ is 
the holonomy $\mathcal{U}_\gamma$.
The loop can be expressed via a (closed) curve ($\gamma_\mathrm{a}$) 
$\gamma_{\mathrm{na}}$ in $\mathscr{M}$ yielding the (adiabatic) nonadiabatic 
holonomy. The embedding $\mathscr{M}$ into $\mathscr{G}_{M,K}$ (blue shaded 
area) does not need to be the same for adiabatic and nonadiabatic holonomies.}
\end{figure}

From a geometric point of view, the $\hat{\eta}_k^\dagger(t)$ are the 
horizontal lifts of a curve $\gamma$ in the Grassmann manifold 
$\mathscr{G}_{M,K}$ containing $K$-dimensional subspaces spanned by the 
operators $\{\hat{\Psi}_j^\dagger \}_j$ (Fig.~\ref{fig:bundle}). Moreover, it can be easily verified 
that the connection is anti-Hermitian,
$(\mathcal{A})_{jk}^\dagger=-(\mathcal{A})_{kj}$, and transforms as a proper 
gauge potential $\mathcal{A}\mapsto G^{-1}\mathcal{A} G+ G^{-1}\md G$ under a 
unitary mixing of operators $\hat{\Psi}_j^\dagger\mapsto \sum_{j} 
G_{jk}\hat{\Psi}_j^\dagger$, $G\in\mathrm{U}(M-2)$. If we further consider the collection of all loops $\gamma$ in $\mathscr{G}_{M,K}$, 
the set $\mathrm{Hol}(\mathcal{A})=\{\mathcal{U}_\gamma\}_\gamma$ forms the 
holonomy group of the (principal fibre) bundle $\mathscr{V}_{M,K}\to 
\mathscr{G}_{M,K}$, where the Stiefel manifold $\mathscr{V}_{M,K}$ is made up of 
$K$-dimensional orthonormal frames $\{\hat{\Psi}_j^\dagger\}_j$. 
As illustrated in Fig.~\ref{fig:bundle}, at every point $\gamma(t)$ there is a 
fibre on which the Lie group 
$\mathrm{U}(M-2)$ acts. Further properties of 
$\mathrm{Hol}(\mathcal{A})$ follow from standard results on differential 
geometry \cite{Nakahara}. 

\subsection{Adiabatic evolution of the star graph} 
\begin{figure}[h]
	\begin{tikzpicture}
		\node at (0,0) {\includegraphics[width=6.4cm]{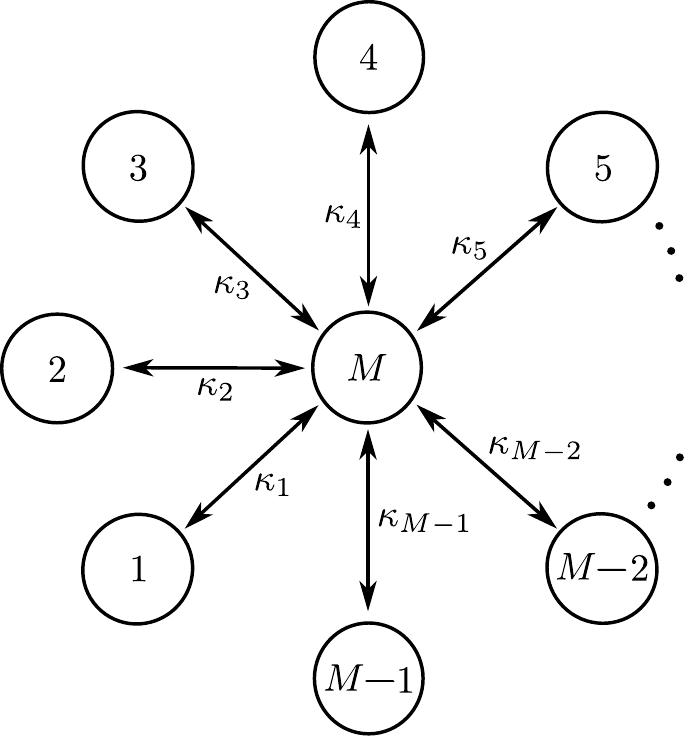}}; 
	\end{tikzpicture}
	\caption{\label{fig:pod} Schematic representation of the star graph corresponding to the $M$-mode system.}
\end{figure}

Consider $M$ bosonic modes being arranged as a star graph (Fig.~\ref{fig:pod}), i.e. its Hamiltonian reads 
\[
\hat{H}(t)=\sum_{k=1}^{M-1}\kappa_{k}(t)\hat{a}_k 
\hat{a}^\dagger_{M}+\kappa_{k}^*(t)\hat{a}_k^\dagger \hat{a}_{M},
\] where the 
couplings $\{\kappa_k\}_k$ act as local coordinates for a $2(M-1)$-dimensional 
control manifold $\mathscr{M}$ which is embedded into
$\mathscr{G}_{M,M-2}$ \cite{Fujii} (Fig.~\ref{fig:bundle}). The system has 
$M-2$ dark modes 
$\hat{D}_j^\dagger(t)=
\kappa_{j+1}(t)\hat{a}_1^\dagger-\kappa_1(t)\hat{a}_{j+1}^\dagger$. 
These operators not only constitute a symmetry, that is 
$[\hat{H}(t),\hat{D}_j^\dagger(t)]=0$, but obey bosonic commutation relations 
$[\hat{D}_j,\hat{D}_k]=[\hat{D}_j^\dagger,\hat{D}_k^\dagger]=0$ and 
$[\hat{D}_j,\hat{D}_k^\dagger]=\delta_{jk}$ after being orthogonalised. Note that,
the demand for dark modes is no limitation at all, if we have eigenmodes with nonzero energy, the dynamical phase can be removed by rescaling the Hamiltonian, so that the condition for a purely geometric evolution is again satisfied.
Note that, an onsite energy $\sigma \hat{a}^\dagger_M\hat{a}_M$ of the central mode leaves the degeneracy structure and dark modes unchanged but only modifies the energy gap $\Delta\varepsilon$ between dark modes and the two remaining eigenmodes, viz. $\Delta\varepsilon \mapsto \sigma/2\pm\sqrt{ \Delta\varepsilon^2+\sigma^2/4}$. Hence, it does not pose a problem to the implementation of the system. However, a distortion of this type occuring in one of the outer modes of the star graph, i.e. $\sigma\hat{a}^\dagger_k\hat{a}_k$ for $k=1,\dots,M-1$, would indeed break the desired degeneracy and would have to be avoided.
The above considerations make it natural to impose a second-quantisation version of the
adiabatic theorem~\cite{Fock} to which the proof can be found in Appendix~\ref{app:sat}, viz.

\textbf{Theorem:} \textit{In the adiabatic limit, any initial operator in 
an subalgebra $\mathscr{A}_{0}$ generated from a (non-)degenerate set of 
eigenmodes will evolve into a final operator lying also in $\mathscr{A}_{0}$ at 
every instance in time.}

Any initial preparation $\hat{D}_{j}^\dagger(0)$ has to reside in the linear span of the modes
$\{\hat{D}_j^\dagger(t)\}_j$ throughout the evolution. The dark modes 
evolve according to the holonomy $\mathcal{U}_\gamma$ governed by the adiabatic 
connection $\left(\mathcal{A}_{\mu}\right)_{jk}=[\hat{D}_k,\partial_\mu 
\hat{D}_j^\dagger]$, $\mu\in\{|\kappa_k|,\mathrm{arg}(\kappa_k)\}_k$ that 
constitutes the operator-valued counterpart of the Wilzeck-Zee connection 
\cite{Wilzeck}.

Similar arguments to those made in Ref.~\cite{Recati} now reveal that  
the connection $\mathcal{A}$ is irreducible for the given system, and hence
$\mathrm{Hol}(\mathcal{A})$ coincides with the entire unitary group 
$\mathrm{U}(M-2)$ (a detailed proof can be found in Appendix~\ref{app:univ}). More specifically, starting the holonomy at an initial 
point $\bm{\kappa}_0=(0,\dots,0,|\kappa|)$ shows that any transformation 
$\sum_{j}(\mathcal{U}_\gamma)_{jk}\hat{D}_j^\dagger(\bm{\kappa}_0)=\sum_{j}
(\mathcal{U}_\gamma)_{jk}\hat{a}_j^\dagger$ can be implemented holonomically by 
designing a suitable loop in $\mathscr{M}$. This means that, due to the 
composition of loops 
$\mathcal{U}_{\prod_j \gamma_j}=\prod_j\mathcal{U}_{\gamma_j}$, any 
linear optical network can be made geometrically robust by supporting it with two auxiliary modes 
$\hat{a}_{M-1}$ and $\hat{a}_{M}$, while adiabatically traversing an 
approximately closed path $\gamma$ in $\mathscr{M}$. 

Of course, this formulation can be related to the standard formalism on 
adiabatic holonomies \cite{Pinske2020}. Excitations of the dark modes produce 
zero-eigenvalue eigenstates (dark states) 
$\ket{\psi_{\bm{n}}}=\prod_j\frac{1}{\sqrt{n_j !}}(\hat{D}_j^\dagger)^{n_j}\ket{\bm{0}}$, 
$\bm{n}\in\mathbb{N}_0^{M-2}$, sharing an adiabatic subspace $\mathscr{H}_0$.
However, these dark modes are not the only ones inducing new dark states. 
The Hamiltonian gives rise in addition to two nondegenerate bright modes 
$\hat{B}_{\pm}^\dagger(t)=(\sqrt{2}\varepsilon)^{-1}(\sum_{j}\kappa_{j}^*(t)\hat{a}_j^\dagger\pm\varepsilon 
\hat{a}_M^\dagger)$, that is,
$[\hat{H},\hat{B}_{\pm}^{\dagger}]=\pm\varepsilon \hat{B}_{\pm}^{\dagger}$, 
where $\varepsilon=(\sum_{j}|\kappa_j|^2)^{1/2}$. The total holonomy of the 
entire system reads 
\[
\mathcal{U}_{0,\gamma}\oplus \me^{\mi\int_{0}^{T}\varepsilon(t)\md t}\mathcal{U}_{+,\gamma}\oplus \me^{-\mi\int_{0}^{T}\varepsilon(t)\md t}
\mathcal{U}_{-,\gamma}.
\]
For photon numbers $N\geq 2$, there exist combinations of $\hat{B}_+^\dagger$ 
and $\hat{B}_-^\dagger$ producing additional dark states. 
The entire 
eigenspace $\mathscr{H}_0=
\mathrm{Span}\{\ket{\psi_{\bm{n}}}\}_{\bm{n}\in\mathbb{N}_0^{M-1}}$ can be 
generated from the subalgebra $\mathscr{A}_0\otimes\mathscr{A}_B\subset\mathscr{A}$ containing sequences of eigenmodes, i.e. 
$\mathscr{H}_0\cong\mathscr{A}_0\otimes\mathscr{A}_B\ket{\bm{0}}$, with $\mathscr{A}_B$ containing excitations of the form $\hat{B}_{+}^\dagger\hat{B}_{-}^\dagger$. More explicitly, any element $\hat{F}$ in 
$\mathscr{A}_0\otimes\mathscr{A}_B$ can be expanded as 
\[
\hat{F}=\!\!\!\sum_{\bm{n}\in\mathbb{N}_0^{M-1}}c_{n_1\dots n_{M-1}}
\frac{\big(\hat{D}_{1}^\dagger\big)^{n_{1}}}{\sqrt{n_1!}}
\frac{\big(\hat{D}_{2}^\dagger\big)^{n_{2}}}{\sqrt{n_2!}}\dots 
\frac{\big(\hat{B}_{+}^\dagger\hat{B}_{-}^\dagger\big)^{n_{M-1}}}{\sqrt{n_{M-1}
!}}
\]
which, by construction, produces an eigenstate with eigenvalue zero, via 
$\hat{F}\ket{\bm{0}}\in\mathscr{H}_0$.  

For illustration, consider a tripod structure 
($M=4$) into which two photons are injected ($N=2$), then there are clearly 
the three dark states 
$\ket{\psi_{20}}=\frac{1}{\sqrt{2}}(\hat{D}_1^\dagger)^2\ket{\bm{0}}$, 
$\ket{\psi_{11}}=\hat{D}_1^\dagger \hat{D}_2^\dagger\ket{\bm{0}}$, 
$\ket{\psi_{02}}=\frac{1}{\sqrt{2}}(\hat{D}_2^\dagger)^2\ket{\bm{0}}$.
Moreover, because of $[\hat{H},\hat{B}_{\pm}^\dagger]=\pm \varepsilon\hat{B}_{\pm}^\dagger$, 
the positive and negative eigenenergies cancel one another out in the case of simultaneous excitation of $\hat{B}_{+}$ and $\hat{B}_{-}$. Therefore,
$\ket{\psi_{+-}}=\hat{B}_+^\dagger \hat{B}_-^\dagger\ket{\bm{0}}$ is another dark 
state which, however, only attains a (scalar) Berry phase while evolving 
adiabatically. Thus, despite the fact that $\{\ket{\psi_j}\}_{j}$ span a 
common eigenspace, $\ket{\psi_{+-}}$ evolves independently. In particular, the corresponding adiabatic evolution in that eigenspace has a block structure in which $\ket{\psi_{+-}}$ does not couple to the other eigenstates~\cite{Pinske2020,Greentree2}.
It can be concluded that demanding the eigenmodes (rather than the eigenstates) 
to evolve adiabatically explains (in contrast to the original formulation \cite{Fock}) why there are eigenstates in $\mathscr{H}_0$ that do not 
couple to the other eigenstates in $\mathscr{H}_0$. This phenomenon was 
observed in Refs.~\cite{Greentree1,Greentree2} but remained, to the best 
of our knowledge, unexplained until now. 

In order to clarify this point further, consider another benchmark Hamiltonian $\hat{H}(t)$. For simplicity, we assume the corresponding eigenmodes to be nondegenerate, that is the modes $\hat{\Psi}_k$ belong to mutually different energies $\varepsilon_k$. We then have the spectral decomposition $\hat{H}=\sum_{k}\varepsilon_k\hat{\Psi}_{k}^\dagger \hat{\Psi}_{k}$, where $\hat{\Psi}_{k}^\dagger \hat{\Psi}_{k}$ acts as a number operator for the $k$th eigenmode. It can be readily checked that $\prod_k\frac{1}{\sqrt{n_k !}}(\hat{\Psi}_k^\dagger)^{n_k}\ket{\bm{0}}$ is an $N$-photon eigenstate with with energy $\sum_{k}\varepsilon_k n_k$ such that $\sum_k n_k=N$.
Interestingly, if the eigenvalues are in such a structure that $N\varepsilon_1=\sum_{k\neq 1}\varepsilon_k n_k$, then the eigenstates $\frac{1}{\sqrt{N!}}(\hat{\Psi}_1^\dagger)^{N}\ket{\bm{0}}$ and $\prod_{k\neq 1}\frac{1}{\sqrt{n_k !}}(\hat{\Psi}_k^\dagger)^{n_k}\ket{\bm{0}}$ both have the same eigenvalue, even though the eigenmodes of the system were nondegenerate.
The original formulation of the adiabatic theorem \cite{Fock} tells us only that (under a slow change of physical parameters) these eigenstates will not couple to states with different eigenvalue. However, the second-quantisation formulation additionally predicts that the states $\frac{1}{\sqrt{N!}}(\hat{\Psi}_1^\dagger)^{N}\ket{\bm{0}}$ and $\prod_{k\neq 1}\frac{1}{\sqrt{n_k !}}(\hat{\Psi}_k^\dagger)^{n_k}\ket{\bm{0}}$ evolve separately from one another as well, because they originate from different eigenmodes. This highlights why the second-quantisation formulation is a stronger version of the adiabatic theorem. In fact, one can use the above argument to construct linear optical networks that give rise to highly-degenerate subspaces~\cite{Pinske2020} via a spectral decomposition with suitable eigenvalue structure.
Finally, note that if only a 
single photon or a coherent state is injected both versions of the adiabatic theorem coincide.

\subsection{Nonadiabatic evolution of the star graph} 
The construction of adiabatic 
holonomies can be formulated analogously for the case where the cyclic evolution is not 
restricted to just the eigenmodes but to a more general collection of modes for 
which dynamical contributions from the Hamiltonian completely disappear. 
We return to the linear optical setting shown in Fig.~\ref{fig:pod}, where $M$ 
bosonic modes are arranged as
a star graph. If we assume that all couplings evolve with the same 
envelope, i.e. $\kappa_k(t)\propto\Omega(t)$, the evolution of dark modes is 
trivially $\hat{\Psi}_j(t)=\hat{D}_j(t)=\hat{D}_j(0)$ for $j=1,\dots,M-2$. Moreover, under these assumptions
it is always possible to find another operator 
$\hat{B}^\dagger=\Omega^{-1}\sum_{j}\kappa_j^*\hat{a}_j^\dagger\in\mathscr{A} 
$ (that is not an eigenmode) such that 
$[\hat{D}_j,\hat{B}^\dagger]=[\hat{a}_M,\hat{B}^\dagger]=0$. The 
time evolution of this operator reads
\begin{equation}
\label{eq:ev}
\hat{\Psi}_{M-1}^\dagger(t)=\me^{\mi\delta(t)}\big(\cos\delta(t)\hat{B}^\dagger(t)
-\mi\sin\delta(t)\hat{a}_{M}^\dagger(t)\big),
\end{equation}
where $\delta(t)=\int_{0}^{t}\Omega(\tau)\md \tau$. One can check that 
$[\hat{\Psi}_j,\hat{\Psi}_k]=[\hat{\Psi}_j^\dagger,\hat{\Psi}_k^\dagger]=0$ and 
$[\hat{\Psi}_j,\hat{\Psi}_k^\dagger]=\delta_{jk}$, hence the orthonormal modes 
$\{\hat{\Psi}_j^\dagger(t)\}_j$ create excitations in a subspace 
$\mathscr{H}_\psi$. Next, we demand $\delta(T)=\pi$ to ensure cyclicity, 
i.e. $\hat{\Psi}_j(T)=\hat{\Psi}_j(0)$ for $j=1,\dots, M-1$. After verifying that
$[\hat{\Psi}_j,[\hat{H},\hat{\Psi}_k^\dagger]]=0$, one can check that 
all conditions for a purely geometric evolution are satisfied. 
The only nonvanishing component of the connection is
$(\mathcal{A}_{t})_{M-1,M-1}=\mi\Omega(t)$, which corresponds to a pure 
gauge. Hence, the nonadiabatic quantum holonomy reads 
$\mathcal{U}_\gamma=\mathrm{diag}(1,\dots,-1)\in\mathrm{U}(M-1)$. 
When replacing the generating operators by the original bosonic modes via a 
change of gauge $\hat{\Psi}_k^\dagger(T)=\sum_j G_{jk} \hat{a}_j^\dagger$, the 
holonomy transforms according to 
$\mathcal{U}_{\gamma}^G(\gamma)=G^{-1}\mathcal{U}_\gamma G$. The unitary 
$\mathcal{U}^{G}_\gamma\in\mathrm{U}(M-1)$ gives rise to noncommutative quantum 
holonomies. In Appendix~\ref{app:univ} it is shown explicitly that suitable manipulation of the couplings $\{\kappa_k\}_k$ allows one to 
design a set of quantum holonomies that generate the entire unitary group $\mathrm{U}(M-1)$. Similar to the 
adiabatic scenario, the holonomy $\mathcal{U}_\gamma^{G}$ can be viewed as a 
scattering matrix describing the unitary mixing of bosonic modes 
$\hat{a}_k^\dagger$ or, equivalently, the superoperator 
$\mathcal{U}_{\gamma}^{G}(\hat{a}^\dagger_k)=\hat{U}^\dagger(\gamma) 
\hat{a}^\dagger_k \hat{U}(\gamma)$. It thus follows that any 
linear optical network can be implemented by means of nonadiabatic geometric 
phases assuming a sufficiently large coupling space, e.g. 
$\mathscr{M}\cong\mathscr{G}_{M,M-1}$. 

As an example we return to the tripod structure 
($M=4$). While the two dark modes evolve trivially in time $\hat{\Psi}_1^\dagger(t)=\hat{D}_{1}^{\dagger}$ and $\hat{\Psi}_2^\dagger(t)=\hat{D}_{2}^{\dagger}$ due to $[\hat{H},\hat{D}_j^\dagger]=0$,
the mode $\hat{B}^\dagger=\Omega^{-1}\big(\kappa_1^*\hat{a}_1^\dagger+\kappa_2^*\hat{a}_2^\dagger+\kappa_3^*\hat{a}_3^\dagger\big)$ evolves into $\hat{\Psi}_3^{\dagger}(t)$ given by Eq.~(\ref{eq:ev}). These modes satisfy the condition for an all-out geometric evolution, i.e. the mixing of modes $\hat{a}_1^\dagger$, $\hat{a}_2^\dagger$, and $\hat{a}_3^\dagger$ is given by a quantum holonomy. 
When two photons ($N=2$) are injected into the optical setup, there are six different states $\hat{\Psi}_j^\dagger\hat{\Psi}_k^\dagger\ket{\bm{0}}$ with $j,k=1,2,3$, spanning a subspace $\mathscr{H}_\psi$ on which a $\mathrm{U}(6)$ holonomy can be implemented. Interestingly, the central mode $\hat{a}_4^\dagger$ of the star graph itself satisfies $[\hat{a}_4,[\hat{H},\hat{a}_4^\dagger]]=0$ and evolves according to an Abelian holonomy $\hat{a}_4^\dagger(\gamma)=\me^{\mi\pi}\hat{a}_4^\dagger(0)$. Hence, the states $\hat{a}_4^\dagger \hat{\Psi}_1^\dagger\ket{\bm{0}}$, $\hat{a}_4^\dagger \hat{\Psi}_2^\dagger\ket{\bm{0}}$, and $\hat{a}_4^\dagger \hat{\Psi}_3^\dagger\ket{\bm{0}}$ span another subspace $\mathscr{H}_{\psi^\prime}$ on which holonomic $\mathrm{U}(3)$ transformations can be performed. This is a general feature of this operator formulation. If there are several subalgebras $\{\mathscr{A}_{\Psi,n}\}_n$, products of modes from different subalgebras will generate a combined subspace on which cyclic evolution leads to a holonomy.

Using standard results~\cite{Browne} we conclude that, when the photonic star graph Hamiltonian is provided with a highly entangled resource state (e.g. a cluster state~\cite{Nielsen}), universal photonic quantum computation is possible using holonomic quantum gates only. This can be done, both for adiabatic and nonadiabatic holonomies, by employing a dual-rail encoding for pairs of modes into which a single photon is injected. In addition, if a linear optical network admits additional symmetries, the geometric phase $\oint_C \mathcal{A}$ might turn into a quantised phase factor, in which case the quantum computation can even be made fully topological~\cite{Simons,Pachos2}.

\section{Extension to a more general framework} 
Formally, one can 
construct an even larger framework. Let $H$ be the generator of 
dynamics belonging to a representation of some dynamical Lie algebra 
$(\mathscr{A},\llbracket\,\cdot\,,\,\cdot\,\rrbracket)$.  The evolution is 
determined by the Lie bracket via the action $\llbracket 
\,\cdot\,, H\rrbracket$. 
Demanding that the bracket vanishes on a well-defined set of solutions 
$\mathscr{K}$ yields a condition for parallel transport in $\mathscr{A}$.
For example, let $\llbracket\,\cdot\,,H\rrbracket_{\mathrm{P}}$ be the flow in 
the phase space $\Gamma$ generated from a classical Hamiltonian 
$H(\bm{q},\bm{p})$ with Darboux coordinates $(\bm{q},\bm{p})$. The 
relevant Lie algebra is the space of all smooth functions $C^\infty(\Gamma)$ 
equipped with the Poisson bracket $\llbracket f,g\rrbracket_{\mathrm{P}}=\sum_j 
(\frac{\partial f}{\partial q^j}\frac{\partial g}{\partial p^j} 
-\frac{\partial g}{\partial q^j}\frac{\partial f}{\partial p^j})$.
The time evolution is then defined by the action $\llbracket 
f,H\rrbracket_{\mathrm{P}}$ for all $f\in C^\infty(\Gamma)$. 
If the system varies periodically with $T$, then under a slow (adiabatic) 
change of external parameters, the explicit time-dependence of $H$ can be 
neglected. Hence, the equations of motion become integrable. For a system with 
$K$ degrees of freedom, it follows that there exists a set of action-angle 
variables $\{\theta_k,J_k\}_k$ such that $\llbracket 
J_k,H\rrbracket_{\mathrm{P}}=0$ for $k=1,\dots, K$, i.e. they are constants of 
motion \cite{Arnold}. Then it follows that $\braket{\dot{J}_k}=T^{-1}\int_{0}^T 
J_k\md t\approx 0$ to satisfactory precision.
By construction, $\braket{\frac{\md}{\md t}}_\mathscr{K}=0$ for $\mathscr{K}$ 
containing the symmetries $\{J_k\}_k$ (they form a subalgebra).
One can find simple mechanical examples. For example, in a 
one-dimensional system with a bounded time-dependent potential 
$V(q,\kappa_\mu(t))$, the relevant subspace contains only a single adiabatic 
invariant $J$. However, adiabatically traversing a closed path 
$\gamma:[0,T]\rightarrow\mathscr{M}$ leads to a change in the generalised 
coordinate $\theta$ 
known as Hannay's angle \cite{Hannay},
$\Delta\theta(\gamma)=\sum_\mu\oint_\gamma \braket{\partial_\mu 
\theta}\md\kappa_\mu$, which depends only on the area enclosed by $\gamma$, thus 
showing a signature of an Abelian holonomy with noncompact symmetry group 
$\mathrm{GL}(1,\mathbb{R})$. 

\section{Conclusions} In this article, we provided a unified framework for 
quantum holonomies based on a holonomic Heisenberg picture. 
We have shown that it provides a remarkable computational advantage for 
bilinear bosonic Hamiltonians where 
the relevant geometric evolution becomes independent of any subspace projection,
and thus enables a description of the holonomy independent of the overall 
photon number. In particular, this means that any linear optical network can 
be constructed using holonomies only. We have shown this explicitly for the 
example of a bosonic star graph Hamiltonian allowing for the generation of 
adiabatic and nonadiabatic quantum holonomies. 
Moreover, we found 
that a stronger version of the adiabatic theorem can be formulated, a 
phenomenon that occurs only in a quantum optical setting.
The parallel transport condition from which these result were derived hints at 
a more general theory that is valid for any dynamical Lie algebra, from which 
the emergence of Hannay's angle follows immediately.  
Our article paves the way to the study of gauge symmetry by quantum optical 
analogies and the realisation of holonomic quantum algorithms using only linear 
optics.  

\acknowledgments
Financial support by the Deutsche Forschungsgemeinschaft (DFG SCHE 612/6-1) 
is gratefully acknowledged.


\appendix

\section{Mode quantisation under geometric constraints}
\label{app:quant}

Consider a collection of $M$ classical 
modes that interact according to a linear optical network. When neglecting any coupling to continuum modes (such as dissipative losses or scattering into the environment), the vector of 
amplitudes $\bm{\alpha}$ transforms according to 
$\bm{\alpha}(T)=\bm{U}\bm{\alpha}(0)$, where $T$ is the 
time it takes to propagate through the optical setup. As such a transformation 
of modes must be unitary, the scattering matrix can be written as
\[
\bm{U}(T)=\mathcal{T}\me^{\mi\int_{0}^{T}\bm{\Phi}(t)\md t},
\]
with $\bm{\Phi}$ being a Hermitian $M\times M$ matrix. The most general 
Hermitian matrix must have components 
$(\bm{\Phi})_{jk}=\kappa_{jk}+\sigma_k\delta_{jk}$, where $\kappa_{jk}=\kappa_{kj}^*$ ($\kappa_{jj}=0$ in this definition) and 
$\sigma_k$ being a real number. As $\bm{\alpha}(t)$ solves the first-order 
differential equation $\partial_t\bm{\alpha}=\mi\bm{\Phi}\bm{\alpha}$, the 
$\kappa_{jk}$ can be viewed as coupling strengths between the modes $j$ and 
$k$, while $\sigma_k$ might be viewed as a self-coupling or a propagation constant. Let us assume that the 
parameter configuration is such that there 
exist $K<M$ orthonormal modes 
$\bm{\Psi}_j(t)=\big(c_{jk}(t)\big)_{k}$ satisfying the condition for 
a purely geometric evolution, that is, for all $j,k=1,\dots, K$ the relation \cite{Anandan} 
\begin{equation}
\label{eq:Hol}
(\bm{\Psi}_j^*)^{\mathrm{T}}\bm{\Phi}\bm{\Psi}_k=\sum_{l,m=1}^{M}c_{jl}^*c_{km}
(\bm{\Phi})_{lm}=0
\end{equation}
holds. Such configurations clearly exist, as the structure of a photonic star graph Hamiltonian
is obtained for $\kappa_{jk}=\kappa_{j}\delta_{jM}$ 
and $\sigma_k=0$ as used in the main article.

Quantisation of this discrete system is carried out by promoting the 
basis vectors to Hilbert space operators $\hat{a}_k^\dagger$. 
Then we have $\bm{\Psi}_k\mapsto \hat{\Psi}_k^\dagger=\sum_{j}c_{jk}\hat{a}_k^\dagger\in\mathscr{A}$. 
Analogously, the Hamiltonian $\hat{H}$ is obtained by comparing the Heisenberg equation 
$\partial_t \hat{a}_k^\dagger=\mi[\hat{H},\hat{a}_k^\dagger]$ to
$\partial_t\bm{\hat{a}}^\dagger=\mi\bm{\Phi}\bm{\hat{a}}^\dagger$ leading to 
\[
\hat{H}(t)=\sum_{j<k}^{M}\kappa_{jk}(t)\hat{a}_j  
\hat{a}^\dagger_{k}+\kappa_{jk}^*(t)\hat{a}_j^\dagger 
\hat{a}_{k}+\sum_{j=1}^{M}\sigma_j(t)\hat{a}_j^\dagger\hat{a}_j.
\]
We now show that, on the level of Hilbert space operators, Eq.~(\ref{eq:Hol}) is 
equivalently represented by the relation 
$[\hat{\Psi}_j,[\hat{H},\hat{\Psi}_k^\dagger]]=0$, thus incorporating the 
condition for parallel transport. Using the properties of the commutator as 
well as the bosonic commutation relations, we arrive at
\begin{widetext}
\begin{equation}
\begin{split}
[\hat{\Psi}_j,[\hat{H},\hat{\Psi}_k^\dagger]]&=\sum_{l,m}c_{jl}^*c_{km}
\Big(\sum_{n<p}\big(\kappa_{np} 
[\hat{a}_l,[\hat{a}_n\hat{a}_p^\dagger,\hat{a}_m^\dagger]]
+\kappa_{np}^*[\hat{a}_l,[\hat{a}_n^\dagger\hat{a}_p,\hat{a}_m^\dagger]]
\big)+\sum_{n}\sigma_n[\hat{a}_l[\hat{a}_n^\dagger\hat{a}_n,\hat{a}_m^\dagger]]
\Big),\\
&=\sum_{l,m}c_{jl}^*c_{km}\Big(\sum_{n<p}
\big(\kappa_{np}\delta_{lp}\delta_{nm}+\kappa_{np}^*\delta_{ln}\delta_{pm}
\big)+\sigma_l\delta_{lm}\Big),\\
&=\sum_{l,m}c_{jl}^*c_{km}\big(\kappa_{lm}+\sigma_l\delta_{lm}\big)
,\\
\end{split}
\end{equation}
\end{widetext}
proving the assertion.

In order to verify that the quantisation procedure indeed leaves Fock states  
with an evolution that is without dynamical contributions, one expects 
that any state $\ket{\psi_{\bm{n}}}$ lying in 
$\mathscr{H}_{\psi}=\mathrm{Span}\{\prod_{j=1}^{K}(\hat{\Psi}_j^\dagger)^{n_j}
/\sqrt{n_j!}\ket{\bm{0}}\,|\,\bm{n}\in\mathbb{N}_{0}^K\}$ satisfies the parallel 
transport condition $\braket{\frac{\md}{\md t}}_{\mathscr{K}}=0$ $\Leftrightarrow$ 
$\braket{\psi_{\bm{n}}|\hat{H}|\psi_{\bm{m}}}=0$ for all sequences 
$\bm{n},\bm{m}\in\mathbb{N}_0^K$. 
To convince oneself that this is indeed the case, we first notice that, if 
both sequences differ in their total photon number,
$\sum_{j}(\bm{n})_j\neq \sum_{j}(\bm{m})_j$, then 
$\braket{\psi_{\bm{n}}|\hat{H}|\psi_{\bm{m}}}=0$ follows immediately, because 
$\hat{H}$ does not alter the total number of photons. Second, the claim is 
obviously true for a single photon, as
$\braket{\psi_{\bm{n}}|\hat{H}|\psi_{\bm{m}}}=\bra{\bm{0}}\hat{\Psi}_{k}\hat{H}
\hat{\Psi}_{j}^\dagger\ket{\bm{0}}=\bra{\bm{0}}[\hat{\Psi}_{k},[\hat{H},\hat{
\Psi}_{j}^\dagger]\ket{\bm{0}}=0$ (we made use of $\hat{H}\ket{\bm{0}}=0$) 
corresponds to the initially assumed parallel transport condition. For two 
photons, note that 
\[
\bra{\bm{0}}\hat{\Psi}_{j}\hat{\Psi}_{k}\hat{H}\hat{\Psi}_{l}^\dagger\hat{\Psi}
_{m}^\dagger\ket{\bm{0}}=\bra{\bm{0}}[\hat{\Psi}_j\hat{\Psi}_k,[\hat{H},\hat{
	\Psi}_l^\dagger \hat{\Psi}_m^\dagger]]\ket{\bm{0}}.
\]
A direct calculation 
reveals that
\begin{widetext}
\[
\begin{split}
\bra{\bm{0}}[\hat{\Psi}_j\hat{\Psi}_k,[\hat{H},\hat{\Psi}_l^\dagger 
\hat{\Psi}_m^\dagger]]\ket{\bm{0}}&=\bra{\bm{0}}\hat{\Psi}_j
\big[\hat{\Psi}_k,[\hat{H},\hat{\Psi}_l^\dagger]\hat{\Psi}_m^\dagger
+\hat{\Psi}_l^\dagger[\hat{ H},\hat{\Psi}_m^\dagger]\big]\ket{\bm{0}},\\
&=\delta_{km}\bra{\bm{0}}\hat{\Psi}_{j}\hat{H}\hat{\Psi}_{l}^\dagger\ket{\bm{0}}
+\delta_{kl}\bra{\bm{0}}\hat{\Psi}_{j}\hat{H}\hat{\Psi}_{m}^\dagger\ket{\bm{0}}=0,\\
\end{split}
\]
\end{widetext}
where we made use of orthogonality relation 
$[\hat{\Psi}_j,\hat{\Psi}_k^\dagger]=\delta_{jk}$ as well as 
$\bra{\bm{0}}\hat{\Psi}_{k}\hat{H}\hat{\Psi}_{j}^\dagger\ket{\bm{0}}=0$ for all $j,k=1,\dots, K$. One can 
continue the argument for higher photon numbers, so that the remainder of the 
proof follows by induction \cite{Zimmermann}.

\section{Proof of the strong adiabatic theorem}
\label{app:sat}

Even though the adiabatic propagation of photon-number states, subject to a 
bilinear Hamiltonian $\hat{H}(t)$, does not violate the original formulation of 
the adiabatic theorem \cite{Fock}, it has become clear that a 
stronger version can be formulated, that is

\textbf{Theorem:} \textit{In the adiabatic limit, any initial operator in an subalgebra 
$\mathscr{A}_{0}(0)$ generated from a (non-)degenerate set of eigenmodes will 
evolve into a final operator lying also in $\mathscr{A}_{0}(t)$ at every instance in time $t$.}

\textit{Proof}. Consider $\hat{H}(t)$ to be the quantum system of 
interest, giving rise to (possibly) degenerate eigenmodes $\hat{\Psi}_{nj}(t)$ 
with eigenvalue $\varepsilon_n(t)$, i.e. 
$[\hat{H},\hat{\Psi}_{nj}^\dagger]=\varepsilon_n\hat{\Psi}_{nj}^\dagger$ at 
every instance $t$. We make the ansatz 
\begin{equation}
 \label{eq:ans}
 \hat{\eta}^\dagger(t)=\sum_{n,j}c_{nj}(t)\hat{\Psi}_{nj}^\dagger(t),
\end{equation}
for the most general bosonic mode of the time-dependent system. When comparing 
the explicit time-derivative of Eq.~(\ref{eq:ans}) with the Heisenberg equation 
of motion 
$\dot{\hat{\eta}}^\dagger=\mi[\hat{H},\hat{\eta}^\dagger]=\mi\sum_{n,j}
\varepsilon_n\hat{\Psi}_{nj}^\dagger$, one arrives at 
\begin{equation}
 \label{eq:deriv}
 \sum_{n,j}\big(\dot{c}_{nj}(t)\hat{\Psi}_{nj}^\dagger(t)+c_{nj}(t)\partial_t\hat{\Psi}_{nj}^\dagger(t)\big)=0,
\end{equation}
where we made use of the Heisenberg equation for the eigenmodes 
$\dot{\hat{\Psi}}_{nj}^\dagger=\mi\varepsilon_n\hat{\Psi}_{nj}
^\dagger+\partial_t \hat{\Psi}_{nj}^\dagger$.
Select the $m$th energy level with $K$-fold degenerate eigenmodes 
$\{\hat{\Psi}_{mk}(t)\}_{k=1}^{K}$, and contract Eq.~(\ref{eq:deriv}) with 
$[\hat{\Psi}_{mk},\,\cdot\,]$. Further, using bosonic commutation relations 
$[\hat{\Psi}_{mk},\hat{\Psi}_{nj}^\dagger]=\delta_{mn}\delta_{kj}$ leads to
\begin{equation}
 \label{eq:evolv}
 \dot{c}_{mk}=-\sum_{n,j}c_{nj}[\hat{\Psi}_{mk},\partial_t\hat{\Psi}_{nj}^\dagger].
\end{equation}
Next, we apply $\partial_t$ to the generalised eigenvalue problem which yields
\[
[\dot{\hat{H}},\hat{\Psi}_{nj}^\dagger]+[\hat{H},\partial_t\hat{\Psi}_{nj}
^\dagger]=\dot{\varepsilon}_n\hat{\Psi}_{nj}^\dagger+\varepsilon_n\partial_t\hat
{\Psi}_{nj}^\dagger,
\]
where we noticed that $\partial_t\hat{H}=\dot{\hat{H}}$.
Contracting this result with $\hat{\Psi}_{mk}$ for $m\neq n$ leaves one with
\begin{equation}
\label{eq:inter}
\varepsilon_n[\hat{\Psi}_{mk},\partial_t\hat{\Psi}_{nj}^\dagger]=[\hat{\Psi}_{mk
},[\dot{\hat{H}},\hat{\Psi}_{nj}^\dagger]]+[\hat{\Psi}_{mk},[\hat{H},
\partial_t\hat{\Psi}_{nj}^\dagger]].
\end{equation}
Using the Jacobi identity it is easy to show that 
$[\hat{\Psi}_{mk},[\hat{H},\partial_t\hat{\Psi}_{nj}^\dagger]]=\varepsilon_m[
\hat{\Psi}_{mk},\partial_t\hat{\Psi}_{nj}^\dagger]$.
With this result, Eq.~(\ref{eq:inter}) can be rewritten in the compact form
\[
[\hat{\Psi}_{mk},\partial_t\hat{\Psi}_{nj}^\dagger]=\frac{[\hat{\Psi}_{mk},[\dot
{\hat{H}},\hat{\Psi}_{nj}^\dagger]]}{\varepsilon_n-\varepsilon_m}.
\]
Inserting the above result into Eq.~(\ref{eq:evolv}) one obtains
\begin{equation}
 \label{eq:evolv2}
\dot{c}_{mk}=-\sum_{j}c_{mj}(\mathcal{A}_t^{(m)})_{jk}-\sum_{n\neq 
m}\sum_{j}c_{nj}\frac{[\hat{\Psi}_{mk},[\dot{\hat{H}},\hat{\Psi}_{nj}^\dagger]]}
{\varepsilon_n-\varepsilon_m},
\end{equation}
with the $K\times K$-matrix $\mathcal{A}^{(m)}$ being the local connection 
one-form for adiabatic parallel transport in the $m$th energy level. Its 
components were defined as 
$(\mathcal{A}_t^{(m)})_{jk}=[\hat{\Psi}_{mk},\partial_t\hat{\Psi}_{mj}^\dagger]$.

An evolution is said to be adiabatic if the Hamiltonian $\hat{H}$ 
changes slowly enough over time $t\in[0,T]$, such that its explicit 
time-dependence can be neglected in the evolution governed by 
Eq.~(\ref{eq:evolv2}). This is clearly the case when
\begin{equation}
 \label{eq:aa}
 \underset{0\leq t\leq 
T}{\mathrm{max}}\big\Vert[\hat{\Psi}_{mk},[\dot{\hat{H}},\hat{\Psi}_{nj}^\dagger
]]\big\Vert\ll \underset{0\leq t\leq 
T}{\mathrm{min}}\big|\varepsilon_n-\varepsilon_m\big|
\end{equation}
giving a validity condition for the adiabatic propagation. On the left-hand 
side of Eq.~(\ref{eq:aa}), we maximise with respect to the induced operator 
norm on $\mathscr{A}$. We further observe that in this adiabatic limit the 
evolution of the components $c_{mk}(t)$ is governed by the system of 
first-order differential equations $\dot{\bm{c}}=\mathcal{A}_t\bm{c}$, with 
$\bm{c}=(c_{mk})_{k=1}^{K}$. In this limit, it becomes evident that the 
dynamical equations for $c_{mk}$ and $c_{nk}$ decouple for $m\neq n$. This means 
that any initial mode $\hat{\eta}^\dagger(0)\in\mathscr{A}_0(0)$ will evolve 
according to $\hat{\eta}^\dagger(T)=\mathcal{T}\me^{\int_0^T\mathcal{A}_t\md 
t}\hat{\eta}^\dagger(0)$ ($\mathcal{T}$ being time ordering) lying in 
$\mathscr{A}_0(T)$. Here, $\mathscr{A}_0(t)$ denotes the subalgebra of 
$\mathscr{A}$ containing analytic functions of eigenmodes 
$\{\hat{\Psi}_{mk}^\dagger(t)\}_{k=1}^{K}$ at time $t$. The decoupling of 
equations for $c_{mk}(t)$ implies further that any operator function 
$\hat{F}(\hat{\Psi}_{mk},\hat{\Psi}_{mk}^\dagger)$, depending solely on the 
eigenmodes of the $m$th level, will reside inside $\mathscr{A}_0(t)$ for all 
$t\in[0,T]$. \hfill$\blacksquare$

\section{Irreducibility of the connection}
\label{app:univ}

Here we show that, for a photonic star graph structure with Hamiltonian
\begin{equation}
 \label{eq:M-pod}
 \hat{H}(t)=\sum_{k=1}^{M-1}\kappa_{k}(t)\hat{a}_k \hat{a}^\dagger_{M}+\kappa_{k}^*(t)\hat{a}_k^\dagger \hat{a}_{M},
\end{equation}
the associated connection $\mathcal{A}$, mediating the parallel transport of a 
bosonic mode, is irreducible. A practical consequence of this statement is that 
it is possible to create any linear optical network by means of holonomies 
only. We give separate proofs for adiabatic and nonadiabatic 
connections, respectively.

\subsection{Adiabatic case}
\label{sapp:univ1}

The Hamiltonian in Eq.~(\ref{eq:M-pod}) possesses $M-2$ (not yet 
orthogonal) dark modes $\hat{D}_j^\dagger(t)= 
\kappa_{j+1}(t)\hat{a}_1^\dagger-\kappa_1(t)\hat{a}_{j+1} ^\dagger$, i.e. 
$[\hat{H},\hat{D}_j^\dagger]=0$ for $j=1,\dots, M-2$ generating a subalgebra 
$\mathscr{A}_0$. After orthogonalisation, these modes satisfy canonical 
commutation relations 
$[\hat{D}_j,\hat{D}_k]=[\hat{D}_j^\dagger,\hat{D}_k^\dagger]=0$, and 
$[\hat{D}_j,\hat{D}_k^\dagger]=\delta_{jk}$.
Under the (strong) adiabatic assumption, any mode from $\mathscr{A}_0(0)$ has to be 
mapped onto an operator in the linear span of
$\{\hat{D}_j^\dagger(T)\}_j$ under time 
evolution according to the holonomy 
$\mathcal{U}_\gamma=\mathcal{T}\mathrm{exp}\int_{0}^{T}\mathcal{A}_t\md t$. 
Here, $\mathcal{T}$ is the time-ordering symbol and 
$(\mathcal{A}_t)_{kj}=[\hat{D}_j,\partial_t\hat{D}_k^\dagger]$ is the local 
connection one-form. Next, let us concentrate on loops of the form
\[
\begin{split}
 \kappa_1&=\kappa\cos\theta\sin\vartheta\me^{\mi\varphi},\\
 \kappa_2&=\kappa\sin\theta\sin\vartheta\me^{\mi\varphi},\\
 \kappa_3&=\kappa\cos\vartheta\\
 \kappa_{4}&=\dots=\kappa_{M-1}=0,\\
\end{split}
\]
where $\theta\in[0,\pi]$ and $\vartheta,\varphi\in[0,2\pi)$.
\begin{figure}[H]
\centering
\begin{tikzpicture}
\node at (0,0) {\includegraphics[width=6cm]{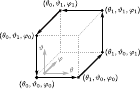}}; 
\end{tikzpicture}
\caption{\label{fig:plaq} Sequence of six steps forming the loop $\gamma(\square)$ in the parameter space $\mathscr{M}$. The hollow dot denotes the starting point $(\theta_0,\vartheta_0,\varphi_0)$.}
\end{figure}
Due to normalisation, the degree of freedom in $\kappa> 0$ can be omitted 
and the remaining coordinates $\{\theta,\vartheta,\varphi\}$ parametrise a 
$3$-dimensional submanifold of $\mathscr{M}$. A simple calculation reveals that 
the fibres over this submanifold are spanned by the dark modes
\[
\begin{split}
 \hat{D}_{1}^{\dagger}&=\sin\theta\hat{a}_1^{\dagger}
 -\cos\theta\hat{a}_2^{\dagger},\\
 \hat{D}_{2}^{\dagger}&=\cos\vartheta\cos\theta\hat{a}_1^{\dagger}
 +\cos\vartheta\sin\theta\hat{a}_2^{\dagger}-\sin\vartheta\me^{\mi\varphi}\hat{a}_{3}^{\dagger},\\
 \hat{D}_{3}^{\dagger}&=\hat{a}_4^\dagger,\\
 &\vdots \\
 \hat{D}_{M-2}^{\dagger}&=\hat{a}_{M-1}^\dagger.\\
\end{split}
\]
The components of the local connection one-form are then computed as
\[
\mathcal{A}_{\theta}=\begin{bmatrix}
0&\cos\vartheta\\
-\cos\vartheta&0\\
\end{bmatrix},\quad \mathcal{A}_\vartheta=0,\quad 
\mathcal{A}_{\varphi}=\begin{bmatrix}
0&0\\
0&\mi\sin^2\vartheta\\
\end{bmatrix}.
\]
Note that we have indeed a non-Abelian gauge potential at our disposal, i.e. 
$[\mathcal{A}_{\theta},\mathcal{A}_{\varphi}]\neq 0$. Now, path ordering in the 
relevant submanifold can be satisfied by traversing a plaquette $\square$ along 
the coordinate lines. To be specific, let us choose the loop $\gamma(\square)$ as depicted in Fig.~\ref{fig:plaq}.
A direct integration along the Wilson lines yields the holonomy as a 
path-ordered product of matrix exponentials
\begin{widetext}
\begin{equation}
\label{eq:holo}
\begin{split}
\mathcal{U}_{\gamma(\square)}&=\mathrm{exp}\left(\int_{\varphi_1}^{\varphi_0}
\mathcal{A}_\varphi\big|_{\vartheta=\vartheta_1}\md\varphi\right)\mathrm{exp}
\left(\int_{\theta_1}^{\theta_0}\mathcal{A}_\theta\big|_{\vartheta=\vartheta_1}
\md\theta\right)\mathrm{exp}\left(\int_{\varphi_0}^{\varphi_1}\mathcal{A}
_\varphi\big|_{\vartheta=\vartheta_0}\md\varphi\right)\mathrm{exp}\left(\int_{
\theta_0}^{\theta_1}\mathcal{A}_\theta\big|_{\vartheta=\vartheta_0}
\md\theta\right),\\
        &=\begin{bmatrix}
           1 & 0\\
           0 & \me^{-\mi\sin^2\vartheta_1\Delta\varphi}\\
          \end{bmatrix}\begin{bmatrix}
           \cos\phi_1 & -\sin\phi_1\\
           \sin\phi_1 & \cos\phi_1\\
          \end{bmatrix}\begin{bmatrix}
           1 & 0\\
           0 & \me^{\mi\sin^2\vartheta_0\Delta\varphi}\\
          \end{bmatrix}\begin{bmatrix}
           \cos\phi_0 & \sin\phi_0\\
           -\sin\phi_0 & \cos\phi_0\\
          \end{bmatrix}
,\\
 \end{split}
\end{equation}
\end{widetext}
where the second equality can be verified by inserting the definitions 
$\Delta\varphi=\varphi_1-\varphi_0$ and 
$\phi_i=\cos\vartheta_i(\theta_1-\theta_0)$ for $i=0,1$. 
The transformation (\ref{eq:holo}) is nothing other than the decomposition of
a general $2\times 2$ unitary matrix, and any element of $\mathrm{U}(2)$ can be 
implemented by traversing a corresponding plaquette.
Physically, this 
means that any optical two-mode transformation between $\hat{a}_1^\dagger$ and 
$\hat{a}_2^\dagger$ [starting at 
$(\theta_0,\vartheta_0,\varphi_0)=(\pi,\pi,0)$] can be 
performed by traversing a closed loop in the 
submanifold $(\theta,\vartheta,\varphi)$. Due to the symmetry of the 
star graph structure, there is no preferred pair of outer waveguides 
$(\hat{a}_j^\dagger,\hat{a}_k^\dagger)_{j\neq k}$ for $j,k=1,\dots, M-2$. 
Hence, traversing closed loops in the corresponding submanifolds of 
$\mathscr{M}$ generates any $\mathrm{U}(2)$ transformation between two arbitrary 
modes. From the point of view of differential geometry, this just corresponds 
to the statement that the holonomy group is independent of the chosen starting 
point \cite{Nakahara}. It is well known that any $\mathrm{U}(M-2)$-mixing 
of bosonic modes $\hat{D}_j^\dagger(0)=\hat{a}_j^\dagger$ ($j=1,\dots,M-2$) 
can be obtained from a sequence of $\mathrm{U}(2)$ transformations acting on a 
pair of modes \cite{Reck}. This shows that the connection $\mathcal{A}$ is 
irreducible and any Reck-Zeilinger-like scheme can be implemented by adiabatic 
quantum holonomies utilising the star graph structure. A particular resource-efficient way to implement such $\mathrm{U}(M-2)$ transformations utilises real-valued couplings $\kappa_{j}$ for $j=1,\dots,M-2$ which are comparatively easier to design than complex parameters. It is only necessary to have one complex coupling $\kappa_{M-1}$ in order to have an Irreducible connection. This is due to the reason that in the star graph, ideally, all outer modes are interchangable. This means that a single ancilla mode can be used to mediate a unitary mixing of any pair of modes $(\hat{a}_j^\dagger,\hat{a}_k^\dagger)_{j\neq k}$ for $j,k=1,\dots, M-2$.

\subsection{Nonadiabatic case}
\label{sapp:univ2}

The proof in the previous section is readily extended to apply to the 
nonadiabatic connection one-form associated with a subspace of the star graph 
structure. This subspace is derived from the subalgebra 
$\mathscr{A}_\Psi$ generated by a set of $M-1$ orthonormal modes 
$\{\hat{\Psi}_j(t)\}_{j}$ satisfying the parallel transport condition
\begin{equation}
\label{eq:PT}
 [\hat{\Psi}_j(t),[\hat{H}(t),\hat{\Psi}_k^\dagger(t)]]=0,
\end{equation}
for all $j,k=1,\dots,M-1$ and at every instance $t$ in an interval $[0,T]$. 
Certainly all dark modes satisfy the condition (\ref{eq:PT}), because 
$[\hat{H}(t),\hat{D}_{j}^\dagger(t)]=0$ and we choose 
$\hat{\Psi}_j(t)=\hat{D}_j(t)$ for $j=1,\dots, M-2$.
By the argument of basis completion, it is always possible to find another 
operator $\hat{B}$ such that 
$[\hat{D}_j(t),\hat{B}^\dagger(t)]=[\hat{a}_M(t),\hat{B}^\dagger(t)]=0$.
In order to further simplify the search for the remaining mode 
$\hat{\Psi}_{M-1}(t)$ (this will not be an eigenoperator), we consider that all 
couplings evolve with the same envelope, i.e. $\kappa_{j}(t)= \Omega(t)g_{j}$, 
with $\Omega(t)$ being a real-valued, piecewise continuously differentiable 
function of time and $\{g_j\}_j$ being constant weights such that 
$\sum_{j}|g_{j}|^2=1$.
First, this implies $\hat{\Psi}_j(t)=\hat{D}_j(0)$ for $j=1,\dots, M-2$. Second, 
the time evolution of the operator 
$\hat{B}^\dagger=\sum_{j=1}^{M-1}g_j^*\hat{a}_j^\dagger$, when subjected to 
the Hamiltonian $\hat{H}(t)=\Omega(t)\hat{h}$, can be obtained from the series 
expansion
\[
\begin{split}
 \hat{U}^\dagger(t)\hat{B}^\dagger\hat{U}(t)&=\hat{B}^\dagger+\sum_{n=1}^{\infty}
\frac{(-\mi\delta)^n}{n!}\underbrace{[\hat{h},[\hat{h},\dots[\hat{h},\hat{B}
^\dagger]]]}_{n-\mathrm{times}},\\
&=\hat{B}^\dagger-\mi\delta[\hat{h},\hat{B}
^\dagger]+\frac{(-\mi\delta)^2}{2}[\hat{h},[\hat{h},\hat{B}^\dagger]]\mp\dots,
\end{split}
\]
where we defined the shorthand $\delta(t)=\int_{0}^{t}\Omega(\tau)\md \tau$. 
Making use of $[\hat{h},\hat{B}^\dagger]=\hat{a}_M^\dagger$ and 
$[\hat{h},\hat{a}_M^\dagger]=\hat{B}^\dagger$, we then get
\[
\begin{split}
 \hat{\Psi}_{M-1}^\dagger(t)&=\me^{\mi\delta(t)}\hat{U}^\dagger(t)\hat{B}^\dagger(t)\hat{U
}(t),\\
&=\me^{\mi\delta(t)}\big(\cos\delta(t)\hat{B}
^\dagger(t)-\mi\sin\delta(t)\hat{a}_{M}^\dagger(t)\big),
\end{split}
\]
where the global phase factor $\me^{\mi\delta(t)}$ has been inserted. We see 
that in the nonadiabatic scenario it is also possible that the central mode 
$\hat{a}_M$ can participate throughout the evolution. Finally, one can check 
that the entire set $\{\hat{\Psi}_j(t)\}_{j}$ satisfies Eq.~(\ref{eq:PT}), 
thus ensuring a purely geometric evolution of bosonic modes. Note that,
under the condition $\delta(T)=\pi$, the generating modes return to their 
initial form after period $T$, viz. $\hat{\Psi}_j(0)=\hat{\Psi}_j(T)$ for all 
$j$. The connection $(\mathcal{A})_{kj}=[\hat{\Psi}_j,\md 
\hat{\Psi}_k^\dagger]$, responsible for describing nonadiabatic parallel 
transport, has only a single nonvanishing component, that is 
$(\mathcal{A}_{t})_{M-1,M-1}=\mi\Omega(t)$, which corresponds to a pure gauge.
This has a geometric interpretation. If the connection corresponds to a pure 
gauge, we have a vanishing curvature \cite{Nakahara}. Nonetheless, one 
can still find nontrivial holonomies, as the vanishing curvature is attributed 
to the fact that we chose a single curve $\kappa_{j}(t)\propto\Omega(t)$ to 
generate the holonomy. 
The one-dimensional space represented by the curve always looks locally like a straight line 
having no curvature [one can attach an ($M-1$)-dimensional Cartesian vielbein along the path]. 

From the explicit form of $\mathcal{A}$ we obtain the 
nonadiabatic holonomy 
$\mathcal{U}_\gamma=\mathrm{diag}(1,\dots,-1)\in\mathrm{U}(M-1)$. Looking at the 
diagonal form of $\mathcal{U}_\gamma$, one might ask if any useful 
transformations can be obtained at all. This is indeed the case, and it becomes 
evident when transforming back to the original bosonic modes 
$\hat{a}_i$ ($i=1,\dots,M-1$) via a change of gauge $G(g_j)\in\mathrm{U}(M-1)$, i.e. 
$\mathcal{U}_\gamma^{G}=G^{\dagger}\mathcal{U}_\gamma G$. In order to prove that 
any element in $\mathrm{U}(M-1)$ can be obtained from a suitable sequence of 
loops $\prod_j \gamma_j$ in $\mathscr{G}_{M,M-1}$, we first set 
\[
\begin{split}
 \kappa_1(t)&=\Omega(t)\sin\left(\theta/2\right)\me^{\mi\varphi},\\
 \kappa_2(t)&=\Omega(t)\cos\left(\theta/2\right),\\
 \kappa_3(t)&=\dots=\kappa_{M-1}(t)=0,
\end{split}
\]
where $(\theta,\varphi)$ are constant parameter angles determining the unitary 
of choice. In this case, the relevant operators are 
\[
\begin{split}
\hat{\Psi}_{1}^\dagger(t)&=\sin(\theta/2)\me^{\mi\varphi}
\hat{a}_{2}^\dagger-\cos(\theta/2)\hat{a}_{1}^\dagger,\\
\hat{\Psi}_{2}^\dagger(t)&=\hat{a}_3^\dagger,\\
&\vdots\\
\hat{\Psi}_{M-2}^\dagger(t)&=\hat{a}_{M-2}^\dagger,\\
\hat{\Psi}_{M-1}^\dagger(t)&=\me^{\mi\delta(t)}\big(\cos\delta(t)
\hat{B}^\dagger-\mi\sin\delta(t)\hat{a}_M^\dagger\big),\\
\end{split}
\]
where
\[
\hat{B}^\dagger=\sin(\theta/2)\me^{-\mi\varphi}\hat{a}_{1}^\dagger
+\cos(\theta/2)\hat{a}_{2}^\dagger,
\]
for the given configuration. The general case can be constructed along similar 
lines but is not necessary because of the following argument.
Under cyclic evolution $\delta(T)=\pi$ the (operator-valued) holonomy becomes
\begin{equation}
 \label{eq:BST}
 \begin{pmatrix}
 \hat{a}_{1}^\dagger(T)\\
 \hat{a}_{2}^\dagger(T)\\
\end{pmatrix}=\begin{pmatrix}
 \cos\theta&-\me^{-\mi\varphi}\sin\theta\\
 -\me^{\mi\varphi}\sin\theta&-\cos\theta\\
\end{pmatrix}\begin{pmatrix}
 \hat{a}_{1}^\dagger(0)\\
 \hat{a}_{2}^\dagger(0)\\
\end{pmatrix}.
\end{equation}
As was noted in Ref.~\cite{Sjoeqvist} for a fermionic system, any matrix 
$\mathcal{U}_\gamma\in\mathrm{U}(2)$ can be realised via a composition of two 
suitable loops $\gamma_{1}$ and $\gamma_2$ in the Grassmann manifold 
$\mathscr{G}_{M,2}$. Note that, in contrast to an adiabatic evolution, here the 
loops $\gamma_1$ and $\gamma_2$ might correspond to open (or even trivial) paths 
in $\mathscr{M}$.
Finally, following the same argument as for the adiabatic case, the holonomy 
group $\mathrm{Hol}(\mathcal{A})$ is independent of the chosen submanifold in 
the coupling space $\mathscr{M}$. Hence, the $\mathrm{U}(2)$ transformations can 
be applied to any pair of modes $(\hat{a}_i,\hat{a}_j)_{j\neq k}$ for all
$j,k=1,\dots M-1$. The fact that we can construct any element of 
$\mathrm{U}(M-1)$ in a fully holonomic fashion is now a direct consequence of the 
argument by Reck \textit{et al.} \cite{Reck}.

\end{document}